\documentstyle[twoside]{article}
\input{epsf}
\input psfig.tex
\begin{document}
%
%
\pagestyle{empty}
\begin{center}
{\Large\bf
CLASSICAL AND QUANTUM ASPECTS\\~\\ OF GRAVITATION AND COSMOLOGY}\\~\\~\\~\\
{\large\bf Proceedings of IAGRG-XVIII\\~\\
February 15 - 17 1996}\\~\\~\\
{ \bf Dedicated to }\\~\\~\\
{\large \bf S. CHANDRASEKHAR}\\~\\
\vskip2cm
{\bf Edited by}\\
\vskip1cm
{\large\bf
G. Date\\~\\
Bala R. Iyer}\\
\vskip6cm
{\large\bf
Institute of Mathematical Sciences Report 117}
\end{center}

\newpage
\setcounter{footnote}{0}
\setcounter{section}{0}
\setcounter{equation}{0}
\setcounter{figure}{0}
\newpage
\def\SHIFT {0.5cm}
\pagestyle{empty}
\begin{center}
{\bf CONTENTS}
\end{center}
\vskip1cm
\begin{tabbing}
~~~~\=~~~~~~~~~~~~~~~~~~~~~~~~~~~~~~~~~~~~~~~~~~~~~~~~~~~~~~~~~~~~~~~~~~~~~~~~~~~~~~~~~~~~~~~~~~~~~~~~~~~~~~~~~~~~~~\=~~~~~\\
\>{\em Foreword}\>\\*[\SHIFT]
\>{\em Scientific organising committee}\>\\*[\SHIFT]
\>{\em Preface}\>\\*[\SHIFT]
1.\>Seeing beauty in the simple and the complex\>\\
\> Chandrasekhar and general relativity\>\\
\>{\em N. Panchapakesan} \>1\\*[\SHIFT]
2.\> On the black hole trail...\>\\
\>A personal journey\>\\
\>{\em C.V. Vishveshwara}\>11\\*[\SHIFT]
3.\>Gravitational waves from inspiralling compact binaries\\ 
\>{\em B. R. Iyer}\> 23\\*[\SHIFT]
4.\>Data Analysis of gravitational wave signals\\
\>from coalescing binaries\\
\> {\em R. Balasubramanian}\>43\\*[\SHIFT]
5.\> Gravitational collapse and cosmic censorship\\
\>{\em T.P. Singh}\>57\\*[\SHIFT]
6.\> Aspects of accretion processes on a rotating black hole\\
\> {\em Sandip Chakrabarti}\>77\\*[\SHIFT]
7.\>Large scale strucure in the universe\\
\>Theory vs observations\\
\>{\em Dipak Munshi}\>93\\*[\SHIFT]
8.\>Some non-linear apsects of cosmological structure formation\\
\>{\em Somnath Bharadwaj}\>105\\*[\SHIFT]
9.\>Radiative corrections to gravitational coupling of neutrinos\\
\>and neutrino oscillations\\
\>{\em G.S. Mohanty}\>109
\end{tabbing}
\newpage
\begin{tabbing}
~~~~\=~~~~~~~~~~~~~~~~~~~~~~~~~~~~~~~~~~~~~~~~~~~~~~~~~~~~~~~~~~~~~~~~~~~~~~~~~~~~~~~~~~~~~~~~~~~~~~~~~~~~~~~~~~~~~~\=~~~~~\\
10.\>Topological defects in cosmology\\
\>{\em Pijush Bhattacharjee}\>115\\*[\SHIFT]
11.\> Generalised Raychaudhuri equations for\\
\>strings and membranes\\
 \>{\em Sayan Kar}\>131\\*[\SHIFT]
12.\> An overview of exact solutions of Einstein's equations\\
\>{\em D.C. Srivatsava}\>143\\*[\SHIFT]
13.\> Quantum gravity and string theory\\
\>{\em J. Maharana}\>155\\*[\SHIFT]
14.\> Eikonal approach to Planck scale physics\\
\>{\em Saurya Das}\>167\\*[\SHIFT]
15.\> Black hole entropy\\
\>{\em Parthasarthi Mitra}\>177\\*[\SHIFT]
16.\>Ashtekar approach to quantum gravity\\
\>{\em G. Date}\>189\\*[\SHIFT]
17.\>Quantum gravity on the computer\\
\>{\em N.D. Hari Dass}\>201\\*[\SHIFT] 
\>{\em List of Contributed Papers}\>\\*[\SHIFT]
\>{\em List of Participants}\>\\*[\SHIFT]
\end{tabbing}

\newpage
\setcounter{footnote}{0}
\setcounter{section}{0}
\setcounter{equation}{0}
\setcounter{figure}{0}
%
%
\pagestyle{empty}
\centerline{{\large{ FOREWORD}}}
\vskip 1.cm
\par
	It gives me great pleasure to write these few words.
\vskip .275cm

	When Prof. Naresh Dadhich suggested the idea that the XVIII th conference
of the IAGRG may be hosted by the Institute of Mathematical Sciences, I
felt it was a welcome opportunity. There was a perception that while
classical general relativity, gravitation, astrophysics, cosmology are
active areas in their own right and as such have been discussed at the
IAGRG meetings in the past, it is perhaps time now to expand the scope of
these meetings to include the quantum gravity and particle physics aspects
as well. Traditionally the general relativity community and the particle
physics community have followed some what non overlapping developments. It
would be mutually beneficial and healthier if both communities can
interact more closely and share their experiences and perceptions. It is
here that IMSc had a significant opportunity to play a role. Some of my
colleagues concurred with this perception and we decided to host the XVIII
the Conference of the IAGRG. 

\vskip .275cm
	Just around the time we took the decision, Prof. S. Chandrasekhar passed
away. In view of the phenomenal contributions of Chandra to General
Relativity and Gravitation, it was but natural to dedicate this meeting to
his memory. IMSc owes a special debt to Chandra as he played a crucial
role in the foundation and the early stages of development of this
Institute. Indeed the birth of the Institute of Mathematical Sciences was
marked by the inaugural lecture by Prof. Chandrasekhar on January 3, 1962
in the lecture halls of Presidency College. 

\vskip .275cm
	This IMSc report reflects the envisaged expanded scope of the IAGRG meeting and
I hope that this trend will continue in the future IAGRG meetings as well. I
also hope that these proceedings will make the frontline developments
accessible to a larger body of researchers in the country particularly to those
from the universities and colleges.

\vskip .275cm
	I may also take this opportunity to thank Drs G. Date and Bala Iyer for 
their efforts as Secretaries in the smooth conduct of the Conference and 
in putting together these proceedings. 

\vskip 0.4cm
\hfill R. Ramachandran 

\newpage
\begin{center}
{\large\bf SCIENTIFIC ORGANISING COMMITTEE}\\
\vskip1cm
\begin{enumerate}
\item Bala R Iyer, RRI, Bangalore (Chairman)
\item Asit Banerji, Jadavpur University, Jadavpur
\item G. Date, IMSc, Madras
\item N.D. Hari Dass, IMSc, Madras
\item Varun Sahni, IUCAA, Pune
\item T.P. Singh, TIFR, Bombay
\item D.C. Srivatsava, Gorakhpur University, Gorakhpur
\end{enumerate}
\end{center}
\vskip2cm
\begin{center}
{\large\bf LOCAL ORGANISING COMMITTEE}\\
\vskip1cm
\begin{enumerate}
\item R. Ramachandran, IMSc, Madras
\item G. Rajasekaran, IMSc, Madras
\item N. D. Hari Dass, IMSc, Madras
\item G. Date, IMSc, Madras
\end{enumerate}
\end{center}


\newpage
\setcounter{footnote}{0}
\setcounter{section}{0}
\setcounter{equation}{0}
\setcounter{figure}{0}
%
%
\pagestyle{empty}
\centerline{\large{PREFACE}}
\vskip 1.0cm
This Institute of Mathematical Sciences Report contains the proceedings of the
XVIII  Conference of the Indian Association for General Relativity and
Gravitation (IAGRG), held during February 15-17, 1996. The conference was
attended by over 50 participants from all over the country, about half of them
being from universities and colleges. The topics range over classical general
relativity, astrophysics, gravity waves, cosmology and quantum aspects of
gravity. The invited talks were intended to give an overview and current status
of research in the respective areas. In addition, there were presentation of
abstracts and theses which were collated in the form of a booklet available to
the participants at the time of conference.

\vskip .3cm
The conference was dedicated to the late Prof. S. Chandrasekhar. 
Prof. N. Panchpakesan kindly agreed to the difficult task of summarising some of Chandra's contributions to General Relativity.
Prof. R. H. Dalitz  who was visiting the
Institute of Mathematical Sciences at the time of conference also kindly
consented to speak on his personal encounters with Chandra. We are thankful to
both of them. 

\vskip .3cm
Prof. C.V. Vishveswara delivered the traditional Vaidya-Raychaudhuri endowment
lecture. We thank him for the very delightful talk. Prof. P.C. Vaidya,
as is usual, made his presense felt throughout the conference. 

\vskip .3cm
Since publication of the proceedings is usually a long drawn process, we felt
that we could bring out the proceedings much faster as an Institute of 
Mathematical Sciences  
Report. In this computer age, with authors doing most of the document
preparation effort, it is relatively easy to put together the proceedings.
Thanks to the various e-print archives available, these proceedings will also
be made available to a much wider set of researchers. We would like to thank
the speakers for giving their manuscripts in tex/latex formats which helped
reduce the editorial work considerably. 

\vskip .3cm
It is a pleasure to acknowledge the encouragement and help we received at
various stages. We thank Prof. R. Ramachandran, Director, Institute of
Mathematical Sciences and Prof. N. Dadhich, President of the IAGRG, for their
active encouragement and participation at all stages of organisation. We thank
Dr. Parthasarthi Majumdar who was a member of the Local organising Committee in
the beginning stages of the conference and who had pointed out that the 
conference 
may be dedicated to Prof. S. Chandrasekhar. Thanks are due to the members of
the Local Organizing Committee and the members of the Scientific Organizing
Committee. The editors thought it would be of some interest to include some
quotes on Chandra and by Chandra. Many of the quotes by Chandra are from the
wonderful biography `CHANDRA' by Kamesh Wali.
\vskip .3cm

We acknowledge the efficient assistance of Mr. Jayaraman, Mr. Sankaran and
their colleagues from the administrative staff; Mr. Sampath and his colleagues
from the guest house staff; the library staff; the computer network of IMSc
and the volunteers from IMSc. 
\vskip .3cm

We  thank G. Manjunath of Raman Research Institute for his patient
help in the preparation of these procedings for final publication. We also
thank Dr. Dipankar Bhattacharya of RRI for advise and assistance in resolving
 vexatious \LaTeX 'nical and Post-Script problems.

\vskip .3cm
We thank the speakers and participants for their active participation. Last but
not the least we thank the Department of Science and Technology, Govt. of India
, the U.G.C and 
the Inter University Center for Astronomy and Astrophysics, Pune  and the
Institute of Mathematical Sciences, Madras for the 
financial support.
\vskip.3cm
\begin{flushright}
G. Date\\
B. R. Iyer 
\end{flushright}
\normalsize
\newpage
\pagestyle{empty}
\mbox{}
\vfill
\begin{quotation}
\sf
I practice style in a very deliberate way. I acquired my style from not
only just reading, for instance, the essays of T.S.Eliot, Virginia Woolf,
and Henry James, but also by paying attention to how they write--how they
construct sentences and divide them into paragraphs. Do they make them
short or long? For example, the idea of just using one sentence for a
paragraph, or of a concluding sentence without subject or object, just a
few words \ldots `so it is' \ldots or some small phrase like that. I
deliberately follow such devices. In fact there is one technique I started
following when I was writing my book  `Radiative Transfer', and I have
followed it since; that is, as you know, in music you repeat periodically
the same phrase in exactly the same form. Very often in my books, when I
have a key idea, and I have to restate it at a later stage, I don't leave
it to chance. I go back and copy exactly what I had written before.
\flushright{--- S. CHANDRASEKHAR}
\end{quotation}

\rm

\newpage
\setcounter{footnote}{0}
\setcounter{section}{0}
\setcounter{equation}{0}
\setcounter{figure}{0}
\pagestyle{empty}
\centerline {\Large {\bf {List of Contributed Papers}} }
\vskip 1.0cm
\begin{enumerate}
\item Static and non static global string\\
A Banerjee, N Banerjee, A Sen 
\item Black hole complementarity and fermions\\
Arundhati Dasgupta 
\item Comments on Yilmaz's theory of gravitation\\
C.S.Unnikrishnan 
\item Cosmic censorship in Tolman Bondi dust collapse\\
C.S.Unnikrishnan 
\item Energy in a gravitational field : Can it make local sense?\\
C.S.Unnikrishnan 
\item Constant Decleration parameter in higher dimension\\
D Panigrahi 
\item The trajectory of particles around cosmic strings\\
G Alagar Ramanujam, K Anandan, G Vasanthakumari 
\item The trajectory of particles in the gravitational field of a tachyon\\
G Alagar Ramanujam, K Meenakshi, H Sridharan 
\item Certain results in the $\lambda$ varying cosmology\\
G Alagar Ramanujam, M Pankajavalli, S Varadharajan 
\item String models in some nontrivial backgrounds\\
G V Vijayagovindan 
\item The orbits of charged particle in an electromagnetic fields on\\ Kerr
background geometry\\
K N Mishra, K Chakraborty 
\item Singularities, hamiltonians and infinite dimensional Lie algebras\\ in
general relativity\\
K.H.Mariwalla 
\item Metric in axially symmetric radiation zone\\
M.D.Patel and R.M.Patel 
\item A Modified Ozer-Taha Type cosmological model\\
Moncy V John, K Babu Joseph 
\item Einstein pseudotensor and total energy of the universe\\
N Banerjee, Somasri Sen 
\item Viscous fluid universe interacting with electromagnetic and zero-mass
scalar fields\\
N Ibotombi Singh 
\item On the relationship between Killing-Yano tensors and electromagnetic\\
fields on curved spaces\\
Ng. Ibohal 
\item Squeezed state representation of black hole radiation\\
P.K.Suresh and V.C.Kuriakose 
\item Gauss map and 2+1 gravity\\
R.Parthasarthy 
\item A cylindrically symmetric stiff fluid solution of Einstein's equations\\
and gravitational collapse\\
Ramesh Tikekar, M C Sabu 
\item How far singularity theorms imply the big-bang singularity? \\
S S Sharma 
\item Distortion of GW signals from binary systems due to presense of
accretion disks\\
Sandip Chakrabarty 
\item Bianchi type-1 Vacuum cosmological model in scale invariant theory \\ of
gravitation \\
Sk. Md. Daud, G. Mohanty 
\item Hermitian Wheeler-Dewitt Operators and the wave function of the
universe\\
Subenoy Chakraborty Nabajit Chakravarty 
\item Spherically symmetric non-static space-time and monopoles\\
Subenoy Chakraborty, Lalit Biswas 
\item Quantum temporal logic, dynamic evolution, and symmetries in the
histories\\ approach to quantum theory\\
Tulsi Dass, Yogesh Joglekar 
\item Can there be a theory of everything in physics?\\
Uma S Sharma 

\end{enumerate}

\newpage
\setcounter{footnote}{0}
\setcounter{section}{0}
\setcounter{equation}{0}
\setcounter{figure}{0}
\setlength{\topmargin}{-.5in}
%
\def\W {14.80cm}
\def\NAMEW {4.0cm}
\def\ADDW {10.0cm}
\def\GAP {0.2cm}
\pagestyle{empty}
\centerline{ {\Large {\bf {List of Participants}} }}
\vskip 1cm 
\noindent
\mbox{ 
\parbox[t]{\W}{ 
\vskip .2cm 
\parbox[t]{\NAMEW}{\centerline {Name} } 
\hspace{0.5cm} 
\parbox[t]{\ADDW}{\centerline {Affiliation} } 
\vspace{\GAP} 
\hrule 
\vskip0.75cm 
\parbox[t]{\NAMEW}{
Alagar Ramanujam G }
\hspace{\GAP}
\parbox[t]{\ADDW}{ 
Principal \& Head, PG Dept. of Phys., NGM College, Pollachi 642 001. }
\vskip .5cm 
\parbox[t]{\NAMEW}{
Balasubramanian R. }
\hspace{\GAP}
\parbox[t]{\ADDW}{ 
IUCAA, Post Bag 4, Ganesh Khind, Pune, 411 007 }
\vskip .5cm 
\parbox[t]{\NAMEW}{
Banerjee Narayan }
\hspace{\GAP}
\parbox[t]{\ADDW}{ 
Dept. of Phys., Jadavpur Univ., Calcutta 700 032. }
\vskip .5cm 
\parbox[t]{\NAMEW}{
Basu Madhumita B. }
\hspace{\GAP}
\parbox[t]{\ADDW}{ 
C/O. DR. S.C. BOSE, 1 Nirmal Ch. St., Calcutta, 700 012. }
\vskip .5cm 
\parbox[t]{\NAMEW}{
Bharadwaj Somnath }
\hspace{\GAP}
\parbox[t]{\ADDW}{ 
Raman Research Inst., C V Raman Avenue, Sadashiva Nagar, Bangalore 560 080. }
\vskip .5cm 
\parbox[t]{\NAMEW}{
Bhattacharya Pijush }
\hspace{\GAP}
\parbox[t]{\ADDW}{ 
Indian Inst. of Astrophysics, Koramangala, Bangalore, 560 034. }
\vskip .5cm 
\parbox[t]{\NAMEW}{
Biswas Lalit }
\hspace{\GAP}
\parbox[t]{\ADDW}{ 
Regional Met. Centre, Met. Dept., Weather Section, 4 Duel Ave., Alipore, Calcutta, 700 027. }
\vskip .5cm 
\parbox[t]{\NAMEW}{
Chakrabarti S. }
\hspace{\GAP}
\parbox[t]{\ADDW}{ 
Theoretical Astrophysics Group, T.I.F.R., Homi Bhabha Rd, Mumbai 400 005. }
\vskip .5cm 
\parbox[t]{\NAMEW}{
Chakraborty Subenoy }
\hspace{\GAP}
\parbox[t]{\ADDW}{ 
Dept. of Math., Jadavpur Univ., Calcutta, 700 032. }
\vskip .5cm 
\parbox[t]{\NAMEW}{
Chakravarty Nabajit }
\hspace{\GAP}
\parbox[t]{\ADDW}{ 
Dept of Math., Jadavpur Univ., Calcutta, 700 032. }
\vskip .5cm 
\parbox[t]{\NAMEW}{
Dadhich Naresh }
\hspace{\GAP}
\parbox[t]{\ADDW}{ 
IUCAA, Post Bag 4, Ganesh Khind, Pune, 411 007 }
\vskip .5cm 
\parbox[t]{\NAMEW}{
Dalitz R.H.}
\hspace{\GAP}
\parbox[t]{\ADDW}{ 
Dept. of Theo. Phys., Oxford Univ., Oxford, UK}
\vskip .5cm 
\parbox[t]{\NAMEW}{
Das Saurya }
\hspace{\GAP}
\parbox[t]{\ADDW}{ 
The Inst. of Mathematical Sciences, Madras, 600 113 }
\vskip .5cm 
\parbox[t]{\NAMEW}{
Dasgupta Arundhati }
\hspace{\GAP}
\parbox[t]{\ADDW}{ 
The Inst. of Mathematical Sciences, Madras, 600 113 }
\vskip .5cm 
\parbox[t]{\NAMEW}{
Date G. }
\hspace{\GAP}
\parbox[t]{\ADDW}{ 
The Inst. of Mathematical Sciences, Madras, 600 113 }
\vskip .5cm 
\parbox[t]{\NAMEW}{
Daud Mahammad }
\hspace{\GAP}
\parbox[t]{\ADDW}{ 
Hariharpur P.O.,Brahmansasan, via Balichak-Midnapur, 721 124. }
\vskip .5cm 
\parbox[t]{\NAMEW}{
Gopalkrishna A.V. }
\hspace{\GAP}
\parbox[t]{\ADDW}{ 
Dept. of Math., I.I.Sc., Bangalore, 560 012 }
\vskip .5cm 
\parbox[t]{\NAMEW}{
Gupta Varsha }
\hspace{\GAP}
\parbox[t]{\ADDW}{ 
Dept. of Phys. \& Astrophysics, Delhi Univ., New Delhi, 110 007. }
\vskip .5cm 
\parbox[t]{\NAMEW}{
Hari Dass N. D. }
\hspace{\GAP}
\parbox[t]{\ADDW}{ 
The Inst. of Mathematical Sciences, Madras, 600 113 }
} } 
\newpage 
\centerline{ {\Large {\bf {List of Participants (Cont...)}} }}
\vskip 1cm 
\noindent
\mbox{ 
\parbox[t]{\W}{ 
\vskip .2cm 
\parbox[t]{\NAMEW}{\centerline {Name} } 
\hspace{0.5cm} 
\parbox[t]{\ADDW}{\centerline {Affiliation} } 
\vspace{\GAP} 
\hrule 
\vskip0.75cm 
\parbox[t]{\NAMEW}{
Ibohal Nagangbam }
\hspace{\GAP}
\parbox[t]{\ADDW}{ 
Dept. of Math., Univ., of Manipur, Imphal, 795 003. }
\vskip .5cm 
\parbox[t]{\NAMEW}{
Iyer B. R. }
\hspace{\GAP}
\parbox[t]{\ADDW}{ 
Raman Research Inst., C V Raman Avenue, Sadashiva Nagar, Bangalore 560 080. }
\vskip .5cm 
\parbox[t]{\NAMEW}{
Joglekar Yogesh N. }
\hspace{\GAP}
\parbox[t]{\ADDW}{ 
B 206, Hall-1, Indian Inst. of Technology, Kanpur, 208 016. }
\vskip .5cm 
\parbox[t]{\NAMEW}{
Jotania Kanti R. }
\hspace{\GAP}
\parbox[t]{\ADDW}{ 
Raman Research Inst., C.V. Raman Avenue, Sadashiva Nagar P.O., Bangalore, 560 080 }
\vskip .5cm 
\parbox[t]{\NAMEW}{
Kar Sayan }
\hspace{\GAP}
\parbox[t]{\ADDW}{ 
Inst. of Physics, Sachivalaya Marg, Bhubaneshwar, 751 005 }
\vskip .5cm 
\parbox[t]{\NAMEW}{
Kuriakose V. C. }
\hspace{\GAP}
\parbox[t]{\ADDW}{ 
Dept. of Phys., Cochin Univ. of Science and Technology, Kochi, 682 022. }
\vskip .5cm 
\parbox[t]{\NAMEW}{
Maharana J. }
\hspace{\GAP}
\parbox[t]{\ADDW}{ 
Inst. of Physics, Sachivalaya Marg, Bhubaneshwar, 751 005 }
\vskip .5cm 
\parbox[t]{\NAMEW}{
Majumdar P. }
\hspace{\GAP}
\parbox[t]{\ADDW}{ 
The Inst. of Mathematical Sciences, Madras, 600 113 }
\vskip .5cm 
\parbox[t]{\NAMEW}{
Mariwala K. }
\hspace{\GAP}
\parbox[t]{\ADDW}{ 
The Inst. of Mathematical Sciences, Madras, 600 113 }
\vskip .5cm 
\parbox[t]{\NAMEW}{
Mishra Kameshwar N. }
\hspace{\GAP}
\parbox[t]{\ADDW}{ 
Dept. of Math.. Bhilai Inst. of Technology, B.I.T., Durg, M.P.}
\vskip .5cm 
\parbox[t]{\NAMEW}{
Mitra P. }
\hspace{\GAP}
\parbox[t]{\ADDW}{ 
Saha Inst. of Nuclear Physics, Block AF, Bidhannagar, Calcutta, 700 064. }
\vskip .5cm 
\parbox[t]{\NAMEW}{
Mohanty S. }
\hspace{\GAP}
\parbox[t]{\ADDW}{ 
Phys. Res. Laboratory, Navarangapura, Ahmedabad, 380 009. }
\vskip .5cm 
\parbox[t]{\NAMEW}{
Moncy V. John }
\hspace{\GAP}
\parbox[t]{\ADDW}{ 
Vilavinal Kozhencherri East P.O. Pathanamthitta, Kerala, 689 641. }
\vskip .5cm 
\parbox[t]{\NAMEW}{
Munshi Deepak }
\hspace{\GAP}
\parbox[t]{\ADDW}{ 
IUCAA, Post Bag 4, Ganesh Khind, Pune, 411 007 }
\vskip .5cm 
\parbox[t]{\NAMEW}{
Panchapakesan N. }
\hspace{\GAP}
\parbox[t]{\ADDW}{ 
Dept. of Phys. Delhi Univ., New Delhi, 110 007. }
\vskip .5cm 
\parbox[t]{\NAMEW}{
Pankajavalli M.}
\hspace{\GAP}
\parbox[t]{\ADDW}{ 
PG Dept of Phys., NGM College, Pollachi, 642 001. }
\vskip .5cm 
\parbox[t]{\NAMEW}{
Parthasarathy R. }
\hspace{\GAP}
\parbox[t]{\ADDW}{ 
The Inst. of Mathematical Sciences, Madras, 600 113 }
\vskip .5cm 
\parbox[t]{\NAMEW}{
Patel M. D. }
\hspace{\GAP}
\parbox[t]{\ADDW}{ 
Dept. of Math., Sardar Patel Univ., Vallabh Vidyanagar, Gujarat, 388 120. }
\vskip .5cm 
\parbox[t]{\NAMEW}{
Prasanna A. R. }
\hspace{\GAP}
\parbox[t]{\ADDW}{ 
Phys. Res. Laboratory, Navarangapura, Ahmedabad, 380 009. }
} } 
\newpage 
\centerline{ {\Large {\bf {List of Participants (Cont...)}} }}
\vskip 1cm 
\noindent
\mbox{ 
\parbox[t]{\W}{ 
\vskip .2cm 
\parbox[t]{\NAMEW}{\centerline {Name} } 
\hspace{0.5cm} 
\parbox[t]{\ADDW}{\centerline {Affiliation} } 
\vspace{\GAP} 
\hrule 
\vskip0.75cm 
\parbox[t]{\NAMEW}{
Raghunathan K. }
\hspace{\GAP}
\parbox[t]{\ADDW}{ 
Dept. of Theo. Phys., Univ. of Madras, Guindy Campus, Madras, 600 025. }
\vskip .5cm 
\parbox[t]{\NAMEW}{
Rajasekaran G. }
\hspace{\GAP}
\parbox[t]{\ADDW}{ 
The Inst. of Mathematical Sciences, Madras, 600 113 }
\vskip .5cm 
\parbox[t]{\NAMEW}{
Rajesh Nayak K. }
\hspace{\GAP}
\parbox[t]{\ADDW}{ 
Indian Inst. of Astrophysics, Koramangala, Bangalore, 560 034. }
\vskip .5cm 
\parbox[t]{\NAMEW}{
Ramachandran R. }
\hspace{\GAP}
\parbox[t]{\ADDW}{ 
The Inst. of Mathematical Sciences, Madras, 600 113 }
\vskip .5cm 
\parbox[t]{\NAMEW}{
Sabu M. C. }
\hspace{\GAP}
\parbox[t]{\ADDW}{ 
Dept. of Math., Sardar Patel Univ., Vallabh Vidyanagar, Gujarat, 388 120. }
\vskip .5cm 
\parbox[t]{\NAMEW}{
Sen Somasri }
\hspace{\GAP}
\parbox[t]{\ADDW}{ 
Rel. \& Cosmo. Research Centre, Dept. of Phys., Jadavpur Univ., Calcutta, 700 032. }
\vskip .5cm 
\parbox[t]{\NAMEW}{
Sen Anjan Ananda }
\hspace{\GAP}
\parbox[t]{\ADDW}{ 
Rel. \& Cosmo. Research Centre Dept. of Phys., Jadavpur Univ., Calcutta, 700 032. }
\vskip .5cm 
\parbox[t]{\NAMEW}{
Sharma Shilendra S. }
\hspace{\GAP}
\parbox[t]{\ADDW}{ 
118/A, Radhaswami Road Houses, Bh/Chankyapuri, Ghatalodiya, Ahmedabad, 380 061. }
\vskip .5cm 
\parbox[t]{\NAMEW}{
Sharma Uma S. }
\hspace{\GAP}
\parbox[t]{\ADDW}{ 
32 Devikrupa Soc., C.T.M. Four Ways Ramol Rd., Ahmedabad, 380 026. }
\vskip .5cm 
\parbox[t]{\NAMEW}{
Singh T.P. }
\hspace{\GAP}
\parbox[t]{\ADDW}{ 
Theoretical Astrophysics Group, T.I.F.R., Homi Bhabha Rd, Mumbai 400 005. }
\vskip .5cm 
\parbox[t]{\NAMEW}{
Srivatsava D.C. }
\hspace{\GAP}
\parbox[t]{\ADDW}{ 
Dept. of Phys., Gorakhpur Univ., Gorakhpur, 273 009. }
\vskip .5cm 
\parbox[t]{\NAMEW}{
Suresh P. K. }
\hspace{\GAP}
\parbox[t]{\ADDW}{ 
Dept. of Phys., Cochin Univ. of Science and Technology, Cochin, 682 022. }
\vskip .5cm 
\parbox[t]{\NAMEW}{
Unnikrishnan C. S. }
\hspace{\GAP}
\parbox[t]{\ADDW}{ 
Gravitational Expts. Group, T.I.F.R., Homi Bhabha Rd, Mumbai 400 005. }
\vskip .5cm 
\parbox[t]{\NAMEW}{
Vaidya P.C. }
\hspace{\GAP}
\parbox[t]{\ADDW}{ 
Dept. of Math., Gujarat Univ., Ahmedabad, 380 009. }
\vskip .5cm 
\parbox[t]{\NAMEW}{
Varadharajan S. }
\hspace{\GAP}
\parbox[t]{\ADDW}{ 
Dept.of Phys., Coll. of Agri. Engg., Pallavaram PO, Lalgudi T.K., Trichy, 621 712. }
\vskip .5cm 
\parbox[t]{\NAMEW}{
Vijayagovindan G. V. }
\hspace{\GAP}
\parbox[t]{\ADDW}{ 
School of Pure and Applied Phys., Mahatma Gandhi Univ., Kottayam, 686 560. }
\vskip .5cm 
\parbox[t]{\NAMEW}{
Vishveshwara C.V. }
\hspace{\GAP}
\parbox[t]{\ADDW}{ 
Indian Inst. of Astrophysics, Koramangala, Bangalore 560 034. }
 } }


\end{document}